\documentclass[preprints,article,accept,moreauthors,pdftex]{Definitions/mdpi}

\firstpage{1} 
\pubvolume{1}
\issuenum{1}
\articlenumber{0}
\pubyear{2021}
\copyrightyear{2020}
\datereceived{} 
\dateaccepted{} 
\datepublished{} 
\hreflink{https://doi.org/} 

\usepackage{color,xspace}

\newcommand{\fig}[1]{Fig.~\ref{fig:#1}}

\newcommand{\citea}[1]{Ref.~\cite{#1}}
\newcommand{\lcdm}[0]{$\Lambda$CDM\xspace}

\Title{Are disks of satellites comprised of tidal dwarf galaxies?}

\TitleCitation{Are disks of satellites comprised of tidal dwarf galaxies?}

\Author{Michal B\'{i}lek$^{1}$\orcidA{}, Ingo Thies$^{2}$, Pavel Kroupa$^{2,3}$\orcidC{}  and Benoit Famaey$^{4}$\orcidD{}}

\AuthorNames{Michal B\'{i}lek, Ingo Thies,  Pavel Kroupa, and Benoit Famaey}

\AuthorCitation{B\'{i}lek, M.; Thies, I.; Kroupa, P.; Famaey, B.}

\address{%
$^{1}$ \quad European Southern Observatory, Karl-Schwarzschild-Strasse 2, 85748 Garching bei M\"unchen, Germany; bilek@asu.cas.cz\\
$^{2}$ \quad Helmholtz-Institut f\" ur Strahlen- und Kernphysik, Nussallee 14-16, 53115 Bonn, Germany\\
$^{3}$ \quad Charles University in Prague, Faculty of Mathematics and Physics, Astronomical Institute, V Hole\v sovi\v ck\' ach 2, 180 00 Prague 8, Czech Republic \\
$^{4}$ \quad   Universit\'e de Strasbourg, CNRS UMR 7550, Observatoire astronomique de Strasbourg, 11 rue de l’Universit\'e, 67000 Strasbourg, France}

\abstract{It was found that satellites of nearby galaxies can form flattened co-rotating structures called disks of satellites or planes of satellites. Their existence is not expected by the current galaxy formation simulations in the standard dark-matter-based cosmology. On the contrary, modified gravity offers a promising alternative: the objects in the disks of satellites are tidal dwarf galaxies, that is small galaxies that form from tidal tails of interacting galaxies. After introducing the topic, we review here our work on simulating the formation of the disks of satellites of the Milky Way and Andromeda galaxies. The initial conditions of the simulation { were} tuned to reproduce the observed positions, velocities and disk orientations of the galaxies. The simulation showed that the galaxies had a close flyby 6.8\,Gyr ago. One of the tidal { tails} produced by the Milky Way was captured by Andromeda. It formed a cloud of particles resembling the disk of satellites at Andromeda by its size, orientation, rotation and mass. A hint of a disk of satellites was formed at the Milky { Way} too. In addition, the encounter induced a warp in the disk of the simulated Milky Way that resembles the real warp by its magnitude and orientation. We present here, for the first time, the proper motions of the members of the disk of satellites of Andromeda predicted by our simulation. Finally, we point out some of the remaining open questions which this hypothesis for the formation of disks of satellites brings up.}


\begin{document}

\section{Introduction}
{ Satellite systems of many nearby galaxies concentrate in flattened co-orbiting structures, the so-called disks of satellites. This contrasts with what is seen in cosmological \lcdm simulations \cite{pawlowski18}. In those, the satellite systems rarely have  ellipticities as high as observed (see, e.g., \cite{schneider12} for a study of the distribution of ellipticities of dark matter halos in simulations; the distribution of satellites is expected to follow the shape of the halo). Moreover, the simulated satellite systems show a significantly lower degree of coherent rotation. This is true both for \lcdm simulations that involve or do not involve baryonic matter \cite{pawlowski15persist}.}

The first example { of a galaxy with a disk of satellites  was found in} our own galaxy, the Milky Way. It was recognized early on that its satellites concentrate toward a plane, called { the Vast Polar Structure}, that is perpendicular to the galactic disk of the Milky Way \cite{lynden-bell76}.  \citea{kroupa05} was the first to point out that this is inconsistent with cosmological simulations. The probability that the distribution of satellites of the Milky Way is isotropic, as the \lcdm simulations predict, came out to be just  0.5\%. It turned up later that most of the satellites also move inside the plane { (their angular momentum vectors are aligned with the satellite plane normal vector)}, most of them orbiting in the same direction \cite{metz08,pawlowski13vpos,pawlowski20,CD+21}. Further satellites are being discovered until now and they still usually lie in the original plane of satellites following the direction of motion of the other satellites \cite{pawlowski14}. The disk of satellites of the Milky Way is traced also by the distribution of stellar streams and young globular clusters \cite{pawlowski12}. The most notable  satellite galaxy that deviates substantially from the other satellites in the terms of location and direction of motion is the Sagittarius Dwarf, which is however { known} to be undergoing a strong tidal interaction with the Milky Way and may have been nudged on its { current orbit by a previous encounter with the Large Magellanic Cloud \cite{zhao98}.} The most massive satellites lie close to the central plane of the disk of satellites, the less massive are more scattered \cite{kroupa12}.

The discovery of this peculiar structure around the Milky Way brought up the question of whether disks of satellites occur also around other galaxies. Can the Milky Way be an exception, one of many thousand cases, as the \lcdm simulations predict? Or is the disk of satellites a look-elsewhere effect? Out of all distributions that that the satellites could have, a disk would not be the only one that would catch our attention. Some random choices of an isotropic distribution could look like a cube, a tetrahedron, an elephant, or any other shape familiar to human mind. The real significance of the fact that the distribution of the satellites of the Milky Way looks interesting for us is difficult to estimate, but is certainly higher than that the satellites would form a plane. These would be some of the main questions that could be clarified by investigating the distribution of satellites around other galaxies. The general difficulty is the faintness of the dwarf galaxies; it is mostly difficult to detect them.

Nevertheless, a disk of satellites was discovered already in the nearest giant galaxy, the Andromeda galaxy \cite{metz07,metz09,ibata13}. Unlike around the Milky Way, only about 50\% of the satellites of Andromeda form a disk. Yet the disk is very thin and statistically significant, such that it cannot be explained by a coincidental clustering of points. It is interesting that the disk of satellites is seen edge-on by an observer in the Milky Way. Proper motions of most of the satellites are not known yet. Still radial velocities suggest that  most  members of the disk of satellites of Andromeda orbit their host in the same direction, just as it is the case for the disk of satellites of the Milky Way \cite{ibata13}.

A satellite plane was found also in the next nearest galaxy, Centaurus~A \cite{tully15,muller16,muller18,muller21}. It is again observed edge-on and most of the satellites co-rotate around  the host. This disk of satellites is inconsistent with both dark-matter-only and hydrodynamical cosmological \lcdm simulations at 0.2\% significance level \cite{muller21}. 

Detecting satellite planes around more distant galaxies becomes more and more difficult. For Andromeda or Centaurus~A we can measure the distances of the satellite precisely enough to reasonably constrain their three-dimensional distribution. One way to detect disks of { satellites} around more distant galaxies is to look for edge-on disks. In this way \citea{heesters21} detected 31 flattened satellite systems around 119 early-type galaxies as a part of the MATLAS survey \cite{duc15, bil20,habas20}. Other searches for satellite planes in distant galaxies use information about the radial velocity. This allowed \citea{paudel21} to detect that the satellites of NGC\,2750 co-rotate around their host. \citea{ibata14} looked for disks of satellites statistically. They investigated a large number of galaxies that host at least two satellites. From the projected positions and radial velocities of the satellites with respect to their hosts, they were able to confirm that satellites indeed often co-rotate around their hosts. 

A phenomenon that might be related to disk of satellites are the planes of dwarf galaxies on the scales of galaxy groups (about 1\,Mpc in diameter). There are two of them in the Local Group  \cite{pawlowski13}. The orientation of the planes is striking. The line connecting the Milky Way and Andromeda lies in the middle of the two planes, being parallel to the line of intersection of the planes. The  disk of satellites of Andromeda lies in the midplane too. In addition, the midplane is nearly parallel to the direction of motion of the Local Group with respect to the cosmic microwave background. Other group-scale dwarf planes were found at the M\,101 galaxy \cite{muller17} and in the M\,81 group \cite{chiboucas13}.

In total, it seems that disks of satellites and dwarf galaxy planes are a common phenomenon. The questions remain of how disks of satellites are  formed and if there is any chance to reconcile their existence with the currently mainstream \lcdm cosmological model. Several explanations of the disks of satellites have been proposed.

\section{Proposals to explain the existence of disks of satellites}
For the Milky Way, it has been proposed that the disk-like distribution of the satellites is just apparent, because a substantial fraction of satellites was discovered by the Sloan Digital Sky Survey, that avoided  sky parts  that are close to the Milky Way equatorial plane. A detailed analysis nevertheless showed that this explanation is not sufficient   \cite{pawlowski16}. Moreover it does not work for the other galaxies hosting satellite planes.

Other proposed explanations of the formation of disks of satellites were based on the way the satellites were accreted by the host. The suggestions included accretion of compact groups of satellites or accretion of  satellites along cosmic filaments. Another hypothesis consisted in an intermediate merger of two galaxies: the satellites of the smaller galaxy formed a disk of satellites around the merger remnant \cite{smith16}. All these mechanisms, however, are automatically included in the cosmological simulations that do not produce disks of satellites in sufficient quantity.

Another explanation for { the existence of disks} of satellites is based on the hypothesis that the satellites are tidal dwarf galaxies. Tidal dwarf galaxies form in tidal tails of interacting galaxies. Their stellar masses can reach $10^{9.5}\,M_\odot$, particularly at high redshifts \cite{ren20}.  Such an origin of the satellites would naturally explain the anisotropic distributions of their position and velocities, as well as some satellites being counter-orbiting \cite{pawlowski11}. It then has to be explained why the internal kinematics of the satellites suggests their high dark-matter content when  tidal dwarf galaxies are not expected to contain dark matter. One option is that the satellites are not in dynamical equilibrium because of tidal interaction with the Milky Way \cite{kroupa97,casas12}. This hypothesis is not perfect either, because many satellites look virialized and their velocity dispersions follow the same baryonic Tully-Fisher relation as isolated galaxies \cite{mcgaugh10}.  In the context of \lcdm cosmology, the satellites of the Milky Way cannot be tidal dwarf galaxies for one more reason. This cosmological model predicts that galaxies with the mass of the Milky Way should have primordial satellites. Their number would depend on the unknown strength of baryonic feedback, but generally the theory expects more satellites than actually observed \cite{klypin99,simon07,sawala14,simpson18}. If all satellites of the Milky Way are tidal dwarf galaxies, then the Milky Way would have to have either no or just a few  primordial satellites. The difficulties of the tidal dwarfs scenario stated above can be removed by assuming cosmologies employing alternative gravity models instead of dark matter, such as the MOND theory \cite{milg83a}. This is will the topic of the rest of this contribution.

The tidal dwarf origin of the disks of satellites  in the Local Group in the context of modified gravity was discussed mostly in  relation with a past encounter of the Milky Way and Andromeda.  It should have induced the formation of tidal arms in the galaxies, that would subsequently collapse into the satellites. This option was supported by the backward analytic calculation of the relative orbit of the Milky Way and Andromeda in MOND  by \citea{zhao13}.  It came out that the galaxies had a very close flyby 7-11\,Gyr ago (depending on uncertain parameters of the problem). This is because the two galaxies are on a nearly radial trajectory. The scenario is suggested also by the observed mutual arrangement of the disks of satellites of the Milky Way and Andromeda. The disks lie nearly in one plane, pointing on each other by their edges, and they rotate in the same sense. It should be pointed out that a similar past close encounter of the Milky Way and Andromeda is not probable in the \lcdm cosmology because when two dark matter halos of comparable masses encounter, they experience strong dynamical friction. The galaxies would have merged on the scale of a few gigayears \cite{mo}. It was found that dynamical friction during major encounters is lower with MOND than with Newtonian gravity and dark matter \cite{tiret08,nipoti08,combtir10,renaud16} (yet the situation is likely opposite for minor mergers \citealp{ciotti06,nipoti08,bil21}). In MOND, moreover, tidal dwarf galaxies form easier \cite{tiret08,renaud16}.

\citea{pawlowski11} demonstrated by Newtonian simulations that galaxy mergers or flybys can lead to the formation of disks of tidal material. Most of this material orbits in the disk, while some parts of it can be counterrotating with respect to the rest. This would explain why the Sculptor dwarf orbits in the disk of satellites of the Milky Way in the opposite sense.  Using simplified $N$-body simulations in MOND, \citea{banik18,banik18b} found that the Milky Way-Andromeda encounter was able to produce not only the disks of satellites at the two galaxies, but also counterparts of the high-velocity dwarf galaxies in the Local Group. Their existence is not expected in the \lcdm cosmology \cite{banik16,banik21}.

It was  not clear from the analytic calculation of \citea{zhao13} how the encounter of the Milky Way and Andromeda affected the galaxies or if such an encounter actually leads to the formation of tidal tails, tidal dwarfs, or disks of satellites. This led us to make a self-consistent $N$-body simulation of the history of the Local Group in MOND \cite[B18 hereafter]{bil18}  (the simplified simulations of \citealp{banik18,banik18b} have not been published at that time).

\section{The simulation of B18}
According to MOND, we encounter the missing mass problem not because of the existence of dark matter, but because our current knowledge of the law of gravity or the law of inertia are not precise. The laws, as we currently know them, are valid only if we deal with systems whose components move with accelerations greater than the constant $a_0 = 1.2\times10^{-10}\,$m\,s$^{-2}$.  If all components of a gravitating system move with accelerations much weaker than $a_0$, then the equations of any MOND theory have to satisfy the space-time scaling symmetry, which reads: if the bodies are allowed to move on some trajectories $r(t)$, then they also have to be allowed to move on trajectories expanded in time and space by a constant factor, $\lambda r(\lambda t)$ \cite{milg09}. Two non-relativistic  \cite{bm84,qumond} and several relativistic full MOND theories have been proposed (see \citealp{famaey12} for a review). In practice, we often predict the dynamics of a system in MOND from the following simple equation \cite{milg83a}, that is exact only for a few specific configurations in the full MOND theories:
\begin{equation}
    \vec{a_N} = \vec{a}\,\mu\left(a/a_0\right).
    \label{eq:mond}
\end{equation}
Here $\vec{a_N}$ denotes the acceleration of the body predicted by the standard Newtonian theory and $\vec{a}$ the actual, i.e. MOND acceleration. The so-called ``interpolation function'' $\mu(x)$ is not known exactly but is required to approach unity for $x\gg1$ (so that the Newtonian dynamics is recovered) and $x$ for $x\ll1$ (so that the space-time scaling invariance emerges). For example,  common choices for $\mu$ are $x/(x+1)$ or $x/\sqrt{x^2+1}$. The equations of MOND are non-linear. This has the external field effect as a consequence: the internal gravity of an object is higher when the object is isolated than when the object resides in an external field (we can imagine for example an isolated dwarf galaxy and a galaxy that has the same distribution of mass but is located next to a massive object, such as a giant galaxy or a galaxy cluster).

Equation~\ref{eq:mond} turned out to well describe dynamics of most galaxies of all morphological types \cite{begeman91,sanders96,deblok98,milg03,milg07,gentile11,milg12,
famaey12,dabringhausen16,lelli17,samur14,bil19}. In relation to the dynamics of the Local Group, it is important that MOND has been shown to be consistent with the dynamics of galaxy groups \cite{milg98,milg18,milg19}. It was possible to construct relativistic extensions of MOND (see the review \citealp{famaey12}), the latest of which agree with the observed speed of gravitational waves and the power spectrum of the cosmic microwave background \cite{skordis19,skordis20}. MOND seems promising to solve some of the long-standing problems of the \lcdm  cosmology, such as the high collision speed of the Bullet Cluster \cite{angus11,katz13,candlish16}, the appearance of the extremely massive  galaxy  cluster ``El Gordo'' at a high redshift \cite{asencio21}, or the unexpectedly low density of the Local Void in the cosmic web and the resulting Hubble tension \cite{haslbauer20}. Signs of the external field effect have been detected too \cite{mcgaugh13a,mcgaugh13b,haghi16,mcgaugh16b,caldwell17,chae20}.

In our simulation, we employed the QUMOND modified gravity theory of MOND \cite{qumond}. In it, the { gravitational} potential $\phi$ is related to the distribution of the density of matter $\rho$ through the equations:
\begin{equation}
    \Delta\phi_\mathrm{N} = 4\pi G\rho, ~~~~~~  \Delta\phi = \vec{\nabla}\cdot\left[ \nu\left(\left|\vec{\nabla} \phi_\mathrm{N}\right|/a_0\right)\vec{\nabla}\phi_\mathrm{N}\right]
\end{equation}
with the boundary conditions $\phi_\mathrm{N}(\vec{r}) \rightarrow 0$ and $\left|\vec{\nabla} \phi(\vec{r})\right| \rightarrow 0$ for $\left|\vec{r}\right|\rightarrow \infty$. Here $\nu$ is the inverse interpolating function defined by the equations $x=y\mu(y), y=x\nu(x)$. We decided for QUMOND for our simulation, because of publicly available software  that is suitable for simulations of galaxy interactions and can work in QUMOND. It is the Phantom of RAMSES code, or PoR\footnote{There is currently another suitable public code, RayMOND \cite{raymond}, that can work not only in QUMOND but also in the AQUAL formulation of MOND \cite{bm84}. It is available at \url{https://www.ifa.uv.cl/sites/graeme/codes.html}.} \cite{por,nagesh21}, that is based on the popular adaptive-mesh-refinement code RAMSES \cite{ramses}.  

\begin{figure}
        \includegraphics[width=10.5 cm]{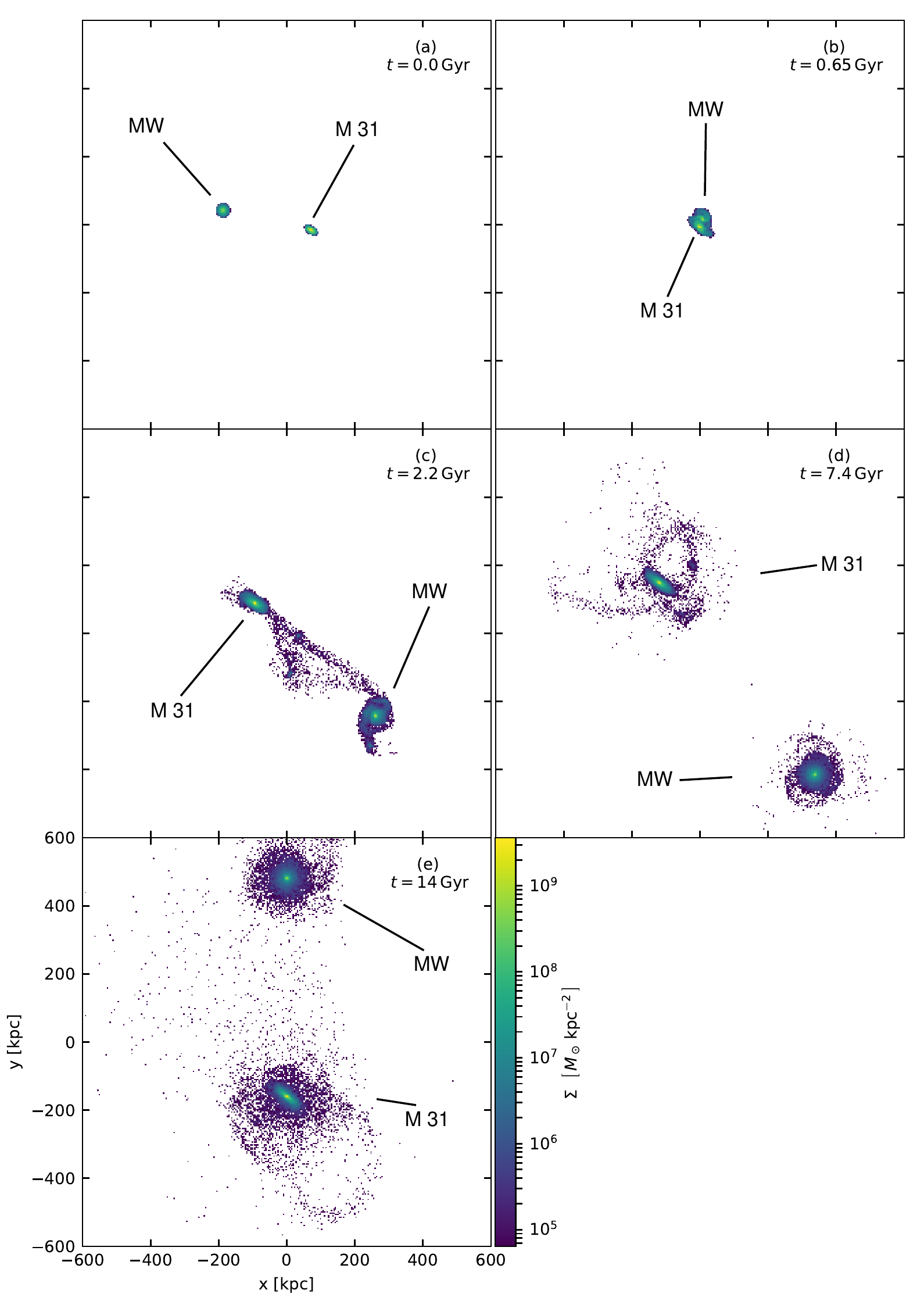}
        \caption{Snapshots from the simulations. (a) Simulation starts. (b) Galaxies are in the first pericenter. (c) Transmission of material from the simulated Milky Way to Andromeda. (d) Time when the observed separation and relative velocities of the galaxies are reproduced, i.e. the current time. (e) Simulation ends. Credit: B\'ilek et al., A\&A, 614, A59, 2018, reproduced with permission \copyright ESO. }
        \label{fig:sim}
\end{figure}

\begin{figure}
        \includegraphics[width=10.5 cm]{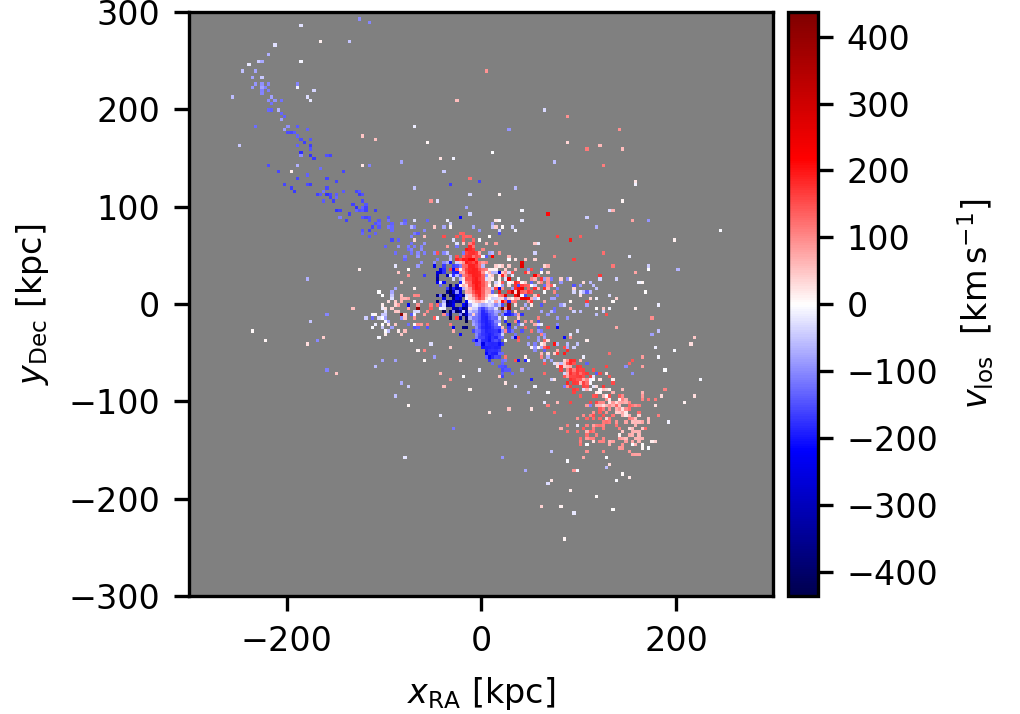}
        \caption{Line-of-sight velocities of particles at the simulated Andromeda galaxy as viewed from the position of the Sun. A  structure resembling the observed satellite plane at the real Andromeda galaxy by its size and rotational pattern. Credit: B\'ilek et al., A\&A, 614, A59, 2018, reproduced with permission \copyright ESO. }
        \label{fig:dos}
\end{figure}

The simulation was done with the spatial resolution of 120\,pc. There were initially $6\times10^4$ particles in the simulated Milky Way  and $1.6\times 10^5$ particles in the simulated Andromeda galaxy. There were only stellar particles in the simulation, no gas. Other galaxies in the Local group were not included in the simulation. The  masses and effective radii of the pre-encounter galaxies were set to be as observed.  The inclinations of the disks of the galaxies with respect to the orbital plane of the galaxies were set as observed too. Once the simulation finished, we verified that  the encounter did not change the parameters of the galaxies substantially. The initial orbital conditions of the encounter were found iteratively. We launched the simulations until we reproduced the observed distance of the galaxies and their radial and transversal velocity. We adopted the observed relative velocity of the galaxies from the measurements by the Hubble Space Telescope \cite{vandermarel12}. The used value was lower but still marginally consistent with the value given by the latest measurements by \citea{salmon21}. In contrast to the analytic model of \citea{zhao13}, our simulation neglected the external field effect arising from the large-scale cosmic structure and the repulsive force between the galaxies caused by cosmic expansion. { The evolution of the simulated interaction is depicted in \fig{sim}.}

The galaxies in the simulation reached their first pericenter 6.8\,Gyr before the current time { (panel (b) of \fig{sim})}, i.e. the time when the observed separation and relative observed velocity of the galaxies were reproduced. This corresponds to a redshift of around 0.8 { in the standard \lcdm cosmology\footnote{Since the \lcdm model is argued to account for many cosmological phenomena, the
cosmology in the final MOND theory, that is yet to be discovered, may resemble that in the \lcdm model.}}. In the pericenter, the centers of the galaxies passed only 24\,kpc apart, when their relative velocity reached about 600\,km\,s$^{-1}$. Such a close encounter induced strong changes in their galactic disks. Both galaxies developed two tidal arms. The arms of Andromeda were rather short and short lived. This was because this galaxy was almost three times more massive than the other and therefore the change of gravitational field in it was not  dramatic. On the other hand, the material in the simulated Milky Way experienced a big change of gravitational potential.   The galaxy produced two tidal arms. The trailing one was ejected toward the Andromeda galaxy { (panel (c) of \fig{sim})}. Its length exceeded 600\,kpc. The gravitational attraction of Andromeda likely contributed to its extreme length. Such a feature would be likely undetectable by the current instruments if it occurred in the real universe, because the material in the tidal tail was very diluted. The simulated tail was partly captured by Andromeda and partly { re-captured} by the Milky Way. The other tidal tail of the Milky Way reached the maximal length of  only about 100\,kpc and was { re-captured} by the Milky Way completely.

At the moment in the simulation that corresponds to the current time { (panel (d) of \fig{sim})}, there was a disk of material around Andromeda. It originated from the tidal arm that was ejected from the Milky Way and was captured by Andromeda. It displayed many characteristics of the observed disk of satellites of Andromeda. It had approximately the correct diameter and pointed on the simulated Milky Way with its edge { ( \fig{dos})}. When viewed from the position of the Sun in the simulation, it showed a smooth gradient of line-of-sight velocity, giving the impression of ordered rotation. It should be mentioned that the structure formed a thin disk only temporarily.  There were even hints of tidal dwarf galaxies in this structure. Our simulation was however probably not suitable for investigating the formation of tidal dwarfs: it was found with Newtonian gravity \cite{wetzstein07}, that if tidal dwarf galaxies form in simulations without gas, then the dwarfs are actually only numerical artifacts that disappear once the simulation has a sufficient mass resolution. Nevertheless, tidal dwarf galaxies in MOND simulations with gas generally form very easily \cite{tiret08,renaud16,thies16}, and therefore it is credible that once our simulation is repeated with gas, there will appear true tidal dwarf galaxies arranged in a disk of satellites. Further satellites in the disk could be descendants of the self-bound  giant gas clumps that are observed in high-redshift galaxies \cite[][e.g.,]{zavadsky17,soto17,fisher17}. Their stellar masses at the relevant redshift reach the values of $10^8\,M_\odot$ \cite{elmegreen09}, commensurable with the Magellanic Clouds.

The tidal arms { captured} by the simulated Milky Way formed a flattened feature too. Its size corresponded roughly to the disk of the Milky Way. The particles, when projected on the sky at the position of the Sun in the simulation, formed streams with coherent motions. This resembled the observed phase-space correlation of the satellites of the Milky Way. 

We noted that the same structures could alternatively be counterparts of the stellar streams observed around the galaxies. The resemblance of the loop-like structures in the simulations to the structures in the halo of Andromeda \cite{ferguson16} was particularly striking. The total mass of the material { captured} to the simulated Andromeda agreed well  with the mass of the streams or the mass of the whole stellar halo of the real Andromeda. The mass of the streams in the simulated Milky Way was too small compared to the observations.

The close encounter of the simulated galaxies  further created a warp in the disk of the Milky Way. Its amplitude and orientation with respect to Andromeda agreed well with observations. Similarly, the encounter made the galactic disks of the simulated galaxies thicker. This could explain the existence of the thick disks of the real galaxies.

Another  piece of evidence supporting the hypothesis of the past encounter of the Milky Way and Andromeda provides the fact that the galaxies possess unusually massive stellar halos. This was found by deep imaging of other galaxies that are similar by stellar mass and Hubble morphology \cite{merritt16}. In addition, \citea{tenjes17} found that the spiral arms in Andromeda cannot have an internal origin. Maybe they were induced by the encounter.

\begin{figure}
        \includegraphics[width=10.5 cm]{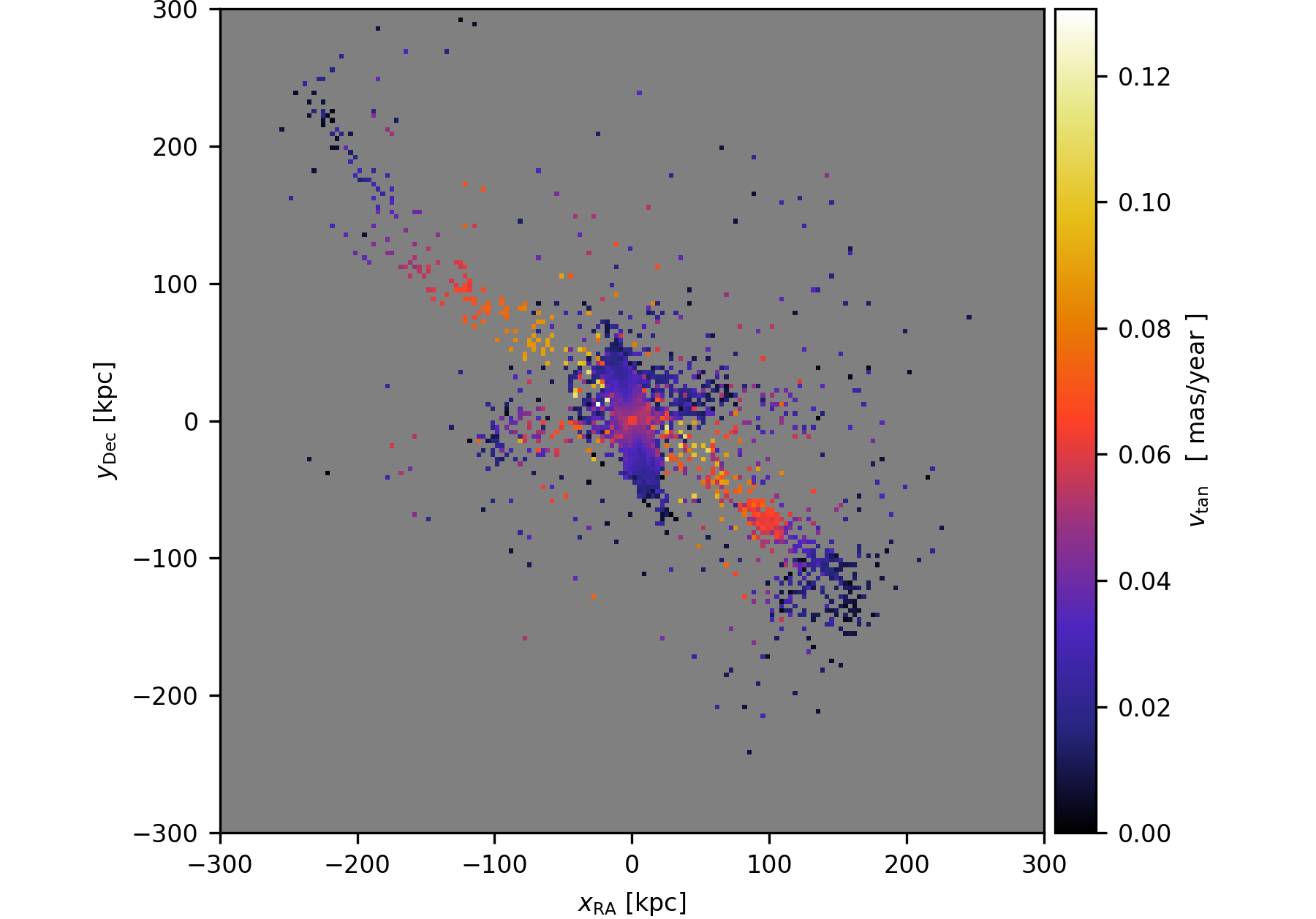}
        \caption{Map of the magnitudes of proper motions of the material at the Andromeda galaxy simulated by B18. The galaxy is viewed from the position of the Sun in the simulation. Proper motion of the whole system was subtracted. }
        \label{fig:mag}
\vspace{6ex}

        \includegraphics[width=10.5 cm]{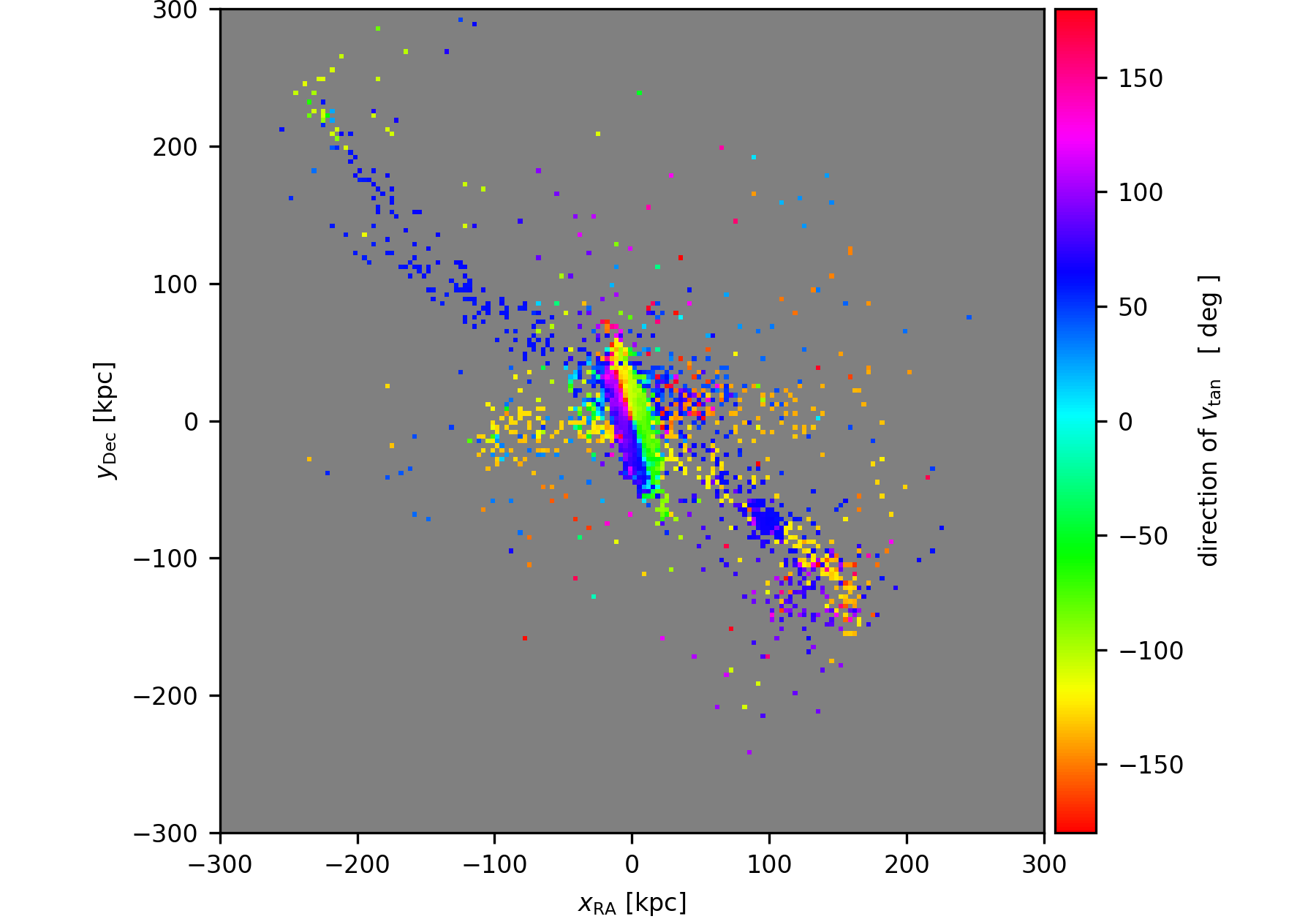}
        \caption{The same as in Figure~\ref{fig:mag} but the direction of the proper motion is shown. The direction is measured counterclockwise from the positive $x$-axis. }
        \label{fig:ang}
\end{figure}

It was argued that if the dwarfs in the disks of satellites are tidal dwarf galaxies, then their properties should be different from primordial dwarf galaxies.  It was found observationally that young tidal dwarf galaxies are more metal rich than other dwarf galaxies because they inherited the metallicities of their giant parents \cite{duc94,duc98,weilbacher03,croxall09}.  \citea{collins15} compared the properties of the members  of the disk of satellites of Andromeda with the satellites of Andromeda that do not lie in the disk. The metallicities were indistinguishable. Nevertheless, as argued by \citea{recchi15}, if tidal dwarf galaxies were created at high redshift, the material of their progenitors would not be metal enriched yet. Such tidal dwarf galaxies would have the same metallicities as other dwarf galaxies. \citea{collins15} compared also effective radii of the satellites inside and outside of the disk of satellites of Andromeda and found no difference. Perhaps surprisingly, tidal dwarf galaxies actually follow the mass-radius relation of usual dwarf galaxies \cite{dabring13,ren20}. \citea{ren20} found that young tidal dwarf galaxies have flatter photometric profiles than standard dwarfs, but this might be because these objects are still assembling. \citea{collins15} neither found a difference in the mass-velocity dispersion relation for the in-plane and off-plane satellites of Andromeda, but this what we expect with MOND \cite{mcgaugh13a,mcgaugh13b}. As a reminder, some of the objects in disk of satellites might not only be genuine tidal dwarf galaxies, but also the descendants of giant gas clumps.

\subsection{Predicted proper motions of the members of the disks of satellites of Andromeda}

Here we present, for the first time,  maps of proper motions of the particles around the simulated Andromeda galaxy. { While many of the particles are not gravitationally bound to each other because of the insufficient resolution of the simulation, the particles are expected to trace the kinematics of tidal dwarf galaxies that an ideal simulation would produce, supposing that the satellites do not interact with each other. Gravitational interactions between the satellites might deflect their orbits.} These figures are intended for comparison of future measurements with proper motions of the members of the disk of satellites of Andromeda. The first measurements of this type have already been presented by \citea{sohn20} who used the Hubble Space Telescope. The satellites are consistent with moving in the plane of the disk of satellites of Andromeda. Other opportunities could provide the next generation of astrometric space missions, such as the proposed Theia satellite \cite{theia}.

Figure~\ref{fig:mag} shows the magnitudes of the proper motions and Figure~\ref{fig:ang} the directions. The axes are chosen to be parallel to the directions of right ascension and declination in an analogue of the equatorial coordinate system in the simulation (see B18 for details). The pixel values correspond to the average velocity of all particles falling in the pixel. The proper motion of the whole system was subtracted. The direction of the proper motion in  Figure~\ref{fig:ang} is measured counter-clockwise from the positive part of the $x$-axis.  The map of radial velocities is presented in B18.  

The figures show that in our simulation, the linear structure moves perpendicularly to its long axis in projection, when viewed from the Sun. This agrees with the fact that the linear structure was found to be a transient feature. In other non-published simulations we ran, most of which were not self-consistent, we found that the { detailed morphology depends  on the details of the encounter such as the initial velocities of the galaxies, their sizes, or on the magnitude of the external field}. Therefore the map in Figure~\ref{fig:mag} should be taken as an order of magnitude estimate and the one in Figure~\ref{fig:ang} should be taken qualitatively: satellites that are close to each other in space should have similar  proper motions. This originates from the high phase-space correlation of the captured tidal tail and therefore this property has to be generic for this formation scenario, regardless of the initial conditions.  { This is the only prediction we can provide now. More detailed predictions would require running a lot of simulations with various values of the unknown parameters.} { Nevertheless, we can already say that the measurements of the proper motions of the two neighboring members of the disk of satellites of Andromeda by \citea{sohn20} showed that the two satellites move in a similar direction and that the magnitude of their proper motion is 0.02 and 0.04 mas\,year$^{-1}$, roughly consistently with the simulation.} { In a \lcdm universe, we still expect some degree of coherent motion of satellites because of accretion of groups of satellites (e.g., \cite{li08}, but see \cite{metz09b}). Nevertheless, the relative velocities of neighboring satellites in \lcdm will most probably be much higher than in the scenario of tidal dwarf galaxies.  The exact comparison is however missing for now, although the related overall performance of the LCDM model needs to be kept in mind \cite{kroupa10,kroupa12,kroupacjp}.}

\section{Open questions, future prospects}

There are several remaining questions regarding the origin of disks of satellites as being tidal dwarf galaxies:

\begin{itemize}
    \item Shortly after the simulation of B18 was published, it was found that here is a population of chemically and kinematically distinct stars in the Milky Way. They were interpreted as a remnant of an accreted dwarf galaxy that got the name Gaia-Enceladus \cite{helmi18} or Gaia-Sausage \cite{belokurov18} and had a mass greater than the Large Magellanic Cloud. Later, evidence for another merger with a galaxy called Sequoia appeared \cite{myeong19}. Could these peculiar stellar populations have been actually  formed by the Milky Way -- Andromeda encounter? In our simulation, the encounter formed only a hint of a disk of satellites around the Milky Way. Could the disk of satellites have  been  formed after the merger with the Gaia-Enceladus/Sausage?
 \item The simulation of B18 contained several modeling simplifications, that can be remedied in the future. It neglected the external field effect and cosmic expansion. As demonstrated by \citea{zhao13}, these have an impact on the time since the encounter and the pericentric distance and velocity. The simulations could be improved by adding gas and star formation, which would help to address the formation and survival of tidal dwarf galaxies during the creation of the disks of satellites. The encounter might have caused a burst of star formation,  whose signature could be looked for observationally. The new simulations should be done taking into account the newest constraint on the proper motion of Andromeda. Another important goal would be to explore a larger space of free parameters. The parameters can include not only the proper motion of Andromeda, but also the initial sizes and gas fractions of the galaxies. Exploration of the parameter space could, for example clarify whether the flyby can produce a disk of satellites of the simulated  Milky Way that has the correct mass and orientation. It would also clarify whether the planes of non-satellite dwarfs in the Local Group could be produced during the flyby.
 \item Is the origin of the planes of non-satellite dwarfs in the Local Group different from the disks of satellites? 
 \item If all satellites of the Milky Way are tidal dwarf galaxies, which is suggested by the positions and velocities of the satellites, is it even possible that the  Milky Way would have no primordial satellites? Pairs of close galaxies in an expanding universe first recede, following the Hubble flow, until their mutual gravity prevails and the galaxies become bound. They will be called a host and a satellite according to their mass. The satellites that have closer apocenters with respect to their hosts became bound at earlier cosmic epochs.  Is it possible that the process of acquisition of satellites, that seems to be  universal for any gravity model, had not worked for the Milky Way?  
 \item As pointed out in B18, if the satellites in the disks are tidal dwarf galaxies, then their globular clusters should not be older than the time since the encounter. For the Local Group that means 7-11\,Gyr. While, for example, some of the GCs of the Fornax dwarf appear older \cite{buonanno98,mackey03,deboer16}, it is generally difficult to exactly determine the ages of stellar populations of this age. I would be desirable to investigate this issue in detail in the future.
 \item Disks of satellites have been detected around all nearby massive galaxies. This suggests that disks of satellites are a common phenomenon. This brings up the question of whether galaxy flybys are frequent enough to explain  all of these disks of satellites by the tidal dwarf scenario. It might be, because a few tens of percent of galaxies at high redshifts are seen to be just undergoing an interaction \cite[e.g.,][]{bridge10,ventou17,ventou19}. A detailed comparison will be desirable once we get a better idea about how frequent disks of satellites are. It will be necessary also to address if all non-merging galaxy flybys that lead to mass transfer lead to the formation of disks of satellites.
\end{itemize}

The formation of disks of satellites will hopefully become clearer once we explore  satellite systems of a higher number of galaxies. For this we need observations that have both a sufficiently low surface brightness limit and a sufficiently high angular resolution. There already are ongoing surveys  whose primary goal is the detection of satellite galaxies of nearby galaxies such as SAGA \cite{geha17} and DGSAT \cite{javanmardi16}. A significant increase of the number of know satellites is expected also from the future large deep sky surveys such as the Euclid satellite or LSST.

\vspace{6pt} 




\funding{This research received no external funding.}

\acknowledgments{MB acknowledges the support by the ESO SSDF grant 21/10.}

\conflictsofinterest{The authors declare no conflict of interest.}




\end{paracol}
\reftitle{References}


\externalbibliography{yes}
\bibliography{citace}

\end{document}